\documentclass[conference]{IEEEtran}
\IEEEoverridecommandlockouts

\usepackage{fancyhdr}
\usepackage{cite}
\usepackage{amsmath,amssymb,amsfonts}
\usepackage{algorithmic}
\usepackage{graphicx}
\usepackage{textcomp}
\usepackage{xcolor}
\usepackage{hyperref} % keep last to avoid package conflicts

\newcommand{\sol}{\textsc{ZenKO}}

\begin{document}

\title{ZenKO: Guided and Iterative Quantum-Classical Method for Molecular Electronic Structure using Sample-Based Quantum Diagonalization}

\author{\IEEEauthorblockN{Aditya Ranjan}
\IEEEauthorblockA{\textit{Northeastern University}\\
\textit{Boston, USA} \\
\texttt{ranjan.ad@northeastern.edu}}
\and
\IEEEauthorblockN{Devesh Tiwari}
\IEEEauthorblockA{\textit{Northeastern University}\\
\textit{Boston, USA} \\
\texttt{d.tiwari@northeastern.edu}}
}

\maketitle

% IEEE copyright notice required on posted accepted versions
% (IEEE Copyright Form, "Author Online Use" / "Electronic Preprints").
\thispagestyle{fancy}
\fancyhf{}
\fancyfoot[C]{\parbox{\textwidth}{\footnotesize \centering
\copyright~2026 IEEE. Personal use of this material is permitted.  Permission from IEEE must be obtained for all other uses, in any current or future media, including reprinting/republishing this material for advertising or promotional purposes, creating new collective works, for resale or redistribution to servers or lists, or reuse of any copyrighted component of this work in other works.}}
\renewcommand{\headrulewidth}{0pt}
\renewcommand{\footrulewidth}{0pt}

\begin{abstract}
The electronic structure of molecules is foundational for scientific advancement, however determining it remains challenging due to the growth of the molecular Hilbert space with system size. Recent approaches address this by using quantum circuits to sample candidate configurations and sample-based quantum diagonalization (SQD) to compute the ground state. We propose \sol{}, a novel gradient-based, iterative SQD method that employs a hybrid quantum-classical approach. \sol{} decomposes configuration selection into discovery, validation, and refinement, enabling targeted subspace expansion and guided exploration. Our evaluation demonstrates that, across $H_2O$, $H_{10}$, and $N_2$ molecules, \sol{} improves ground-state accuracy compared to prior iterative methods, while significantly reducing classical cost.
\end{abstract}


\section{Introduction}
\label{sec:intro}

\noindent\textbf{Problem Background, Significance, and Challenge.}  Many societally important applications, including drug discovery, energy technologies, and materials design, depend on accurately understanding the electronic structure of molecules\cite{apps1,apps2,apps3}. This involves determining the energy levels and quantum states of a molecular system, which in turn govern properties such as stability, bonding, and chemical reactivity.

\vspace{1mm}

Conceptually, the electronic structure problem is formulated in terms of molecular orbitals, which serve as occupancy sites for electrons~\cite{orbital}. Each basis state corresponds to one valid electron configuration across these molecular orbitals, and the full solution space consists of all such configurations. \textit{The goal is to determine the lowest-energy quantum state, or \textbf{ground state}--which plays a central role in determining molecular stability, bonding, and reactivity--and is expressed as a superposition of multiple electron configurations~\cite{ground_state, orbitals2}}.

The number of valid electron configurations grows combinatorially with molecular system size~\cite{size1, size2, size3}. The molecular Hamiltonian needs to be constructed over this large configuration space, and solving the problem requires computing its eigenvalues and eigenstates.  As a result, exact diagonalization becomes computationally expensive on classical HPC systems~\cite{intractibility1, intractibility2}. Therefore, traditional quantum chemistry relies on approximation methods, such as the Hartree–Fock~\cite{hf}

\noindent and Coupled-cluster methods~\cite{ccsd}.  These techniques reduce computational cost by making simplifying assumptions on the wavefunction and its correlations. While effective for some problems, they are inadequate for molecular systems with different properties and complexities~\cite{ccsd_lim1, ccsd_lim2}.

\vspace{2mm}

\noindent\textbf{Hybrid Quantum--HPC Approach.} Quantum algorithms offer an alternate path to reducing the complexity of electronic structure problems by identifying electron configurations that form the \emph{ground state}~\cite{quantum_chem}. In particular, Sample-based Quantum Diagonalization (SQD)~\cite{qsci, te_sqd} uses \textit{quantum circuits} to sample important configurations, constructs a reduced electron configuration \textit{subspace} on which the Hamiltonian is projected and \textit{diagonalized using classical computation}~\cite{qsci}. 

\vspace{1mm}

This hybrid quantum-classical approach is appealing because it avoids solving the full combinatorial problem classically, while retaining the ability to recover the exact ground state: \textit{if the sampled configuration subspace captures all the configurations in the ground state, the exact solution can be recovered after classical diagonalization.} Unlike fixed-form classical approximations, the accuracy of SQD is therefore governed primarily by the quality of the sampled subspace from the quantum circuit.  However, if the quantum circuit samples many irrelevant electron configurations (configurations not contributing to the ground state), classical diagonalization of the reduced problem becomes expensive.

\begin{figure*}
    \centering
    \includegraphics[width=1\linewidth]{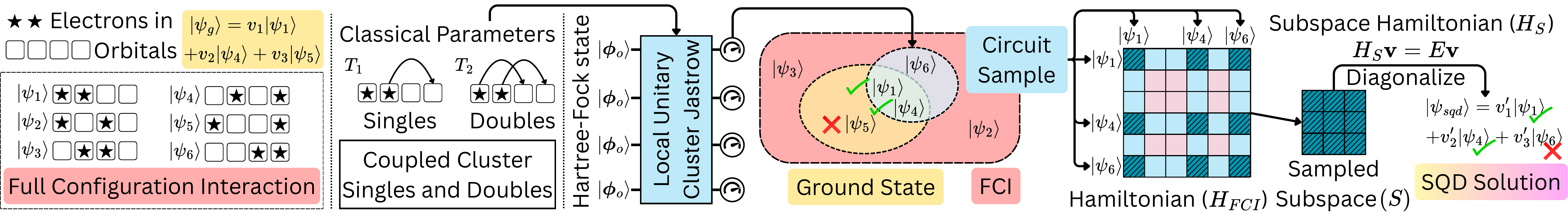}
    \vspace{-7mm}
    \caption{\textbf{Left:} Example configuration space for 2 electrons in 4 orbitals ($^4C_2=6$). The exact ground state $|\psi_g\rangle$ is a linear combination of configurations. \textbf{Middle:} CCSD approximates the ground state using single and double excitations of the Hartree--Fock reference. \textbf{Right:} SQD samples configurations from a LUCJ circuit initialized with CCSD parameters to define a subspace $\mathcal{S}$. The Hamiltonian projected onto $\mathcal{S}$ is diagonalized to obtain the SQD solution $|\psi_{\mathrm{sqd}}\rangle$.
    \vspace{-5mm}}
    \label{fig:background}
\end{figure*}

\vspace{2mm}
\noindent\textbf{Limitations of Existing Methods.} \textit{Existing approaches attempt to address this challenge using classically informed initialization of quantum circuits, time-evolved states, or augmented configuration generation~\cite{sqd1, sqd2, te_sqd, ext_sqd}, but these strategies remain largely static or ineffective}: once samples are generated, there is limited ability to iterate and refine the sampling process based on the intermediate results. 

\vspace{1mm}

More recently, iterative hybrid approaches have begun to incorporate feedback from the current solution to improve sampling over time~\cite{shirakawa2025}, highlighting the promise of adaptive quantum-classical workflows. However, they still rely on repeated subspace construction and classical diagonalization; \textit{as a result, finding the balance between classical computation time (diagonalization of the subspace) and solution quality (discovery of ground state and its corresponding energy) remains challenging.} These limitations motivate the need for a more efficient iterative framework that more directly leverages feedback, guides exploration, and reduces reliance on repeated expensive classical computation.

\vspace{2mm}
\noindent\textbf{\sol{}: Key Ideas Behind A New Hybrid Quantum-HPC Solution.}  
\sol{} is a novel quantum-classical iterative optimization framework with three key ideas: \textit{guided discovery} of new configurations, \textit{weighted refinement} toward promising parameters, and exploitation of \textit{excitation parameters} to improve \sol{}'s efficiency and accuracy.

\sol{}'s discovery process identifies previously unexplored configurations with \textit{guided discovery} to find relevant configurations at relatively lower classical cost, by controlling the size of the discovered configuration set. This is accomplished via Hamiltonian coupling as a proxy for the potential contribution of the configuration to the ground state. \sol{} applies the SQD process to reduce the discovered set to a smaller, more useful subset consisting only of relevant configurations. This subset, along with weights reflecting each configuration’s contribution to the ground state, is used during \textit{weighted refinement} to update the quantum circuit parameters. \sol{} reveals that decoupling the circuit parameters (\textit{excitation parameters}) yields better solution accuracy. 

\vspace{1mm}

In particular, \sol{} uses single-excitation parameters, which govern more local interactions, for refinement, and double-excitation parameters, which capture broader interactions, are used during the discovery phase.

\vspace{2mm}
\noindent\textbf{\sol{}: Evaluation and Results.}  Our evaluation demonstrates that \sol{} improves both sides of the Sample-based Quantum Diagonalization (SQD) tradeoff:  \textit{\sol{} achieves higher ground-state accuracy while substantially lowering classical computing cost -- on a representative, diverse range of molecular systems.} For example,  compared to prior methods, \sol{} reaches near-exact accuracy on $H_{10}$ ($<0.1$ mHa vs.\ 5.0 mHa) and cuts error on $N_2$ from 8.7 to 2.3 mHa, while reducing diagonalization time by 4.2$\times$ on $H_{10}$ and 7.8$\times$ on $N_2$. Importantly, \sol{}'s effectiveness is not isolated to a single setting: it remains consistent across bond-length sweeps, larger hydrogen chains, and under noise.  The \sol{} framework is open-sourced and available at - \href{https://doi.org/10.5281/zenodo.21541111}{https://doi.org/10.5281/zenodo.21541111}.

\section{Background and Motivation}
\label{sec:background}

\noindent \textbf{Electronic Spectrum as a Diagonalization Problem.} A molecular system with $n$ electrons in $o$ orbitals is described by configurations specifying how electrons occupy the orbitals. Each configuration defines a basis state, and the full configuration space consists of all such configurations, with size $^oC_n$ (neglecting spin). As illustrated in Fig.~\ref{fig:background} (left), a system with 2 electrons in 4 orbitals has $^4C_2 = 6$ configurations. However, this number grows combinatorially; for example, $^{40}C_{20} \approx 10^{11}$. The ground state is expressed as a linear combination of configurations in this space. The molecular Hamiltonian is defined on this configuration space, yielding a $^oC_n \times ^oC_n$ matrix whose diagonalization yields the ground and excited states. This scaling renders exact diagonalization impractical. Hartree--Fock (HF) approximates the ground state using a single configuration, leading to self-consistent equations for the orbitals. This approximation has been systematically improved by incorporating excited configurations.

\begin{figure}
    \centering
    \includegraphics[width=1\linewidth]{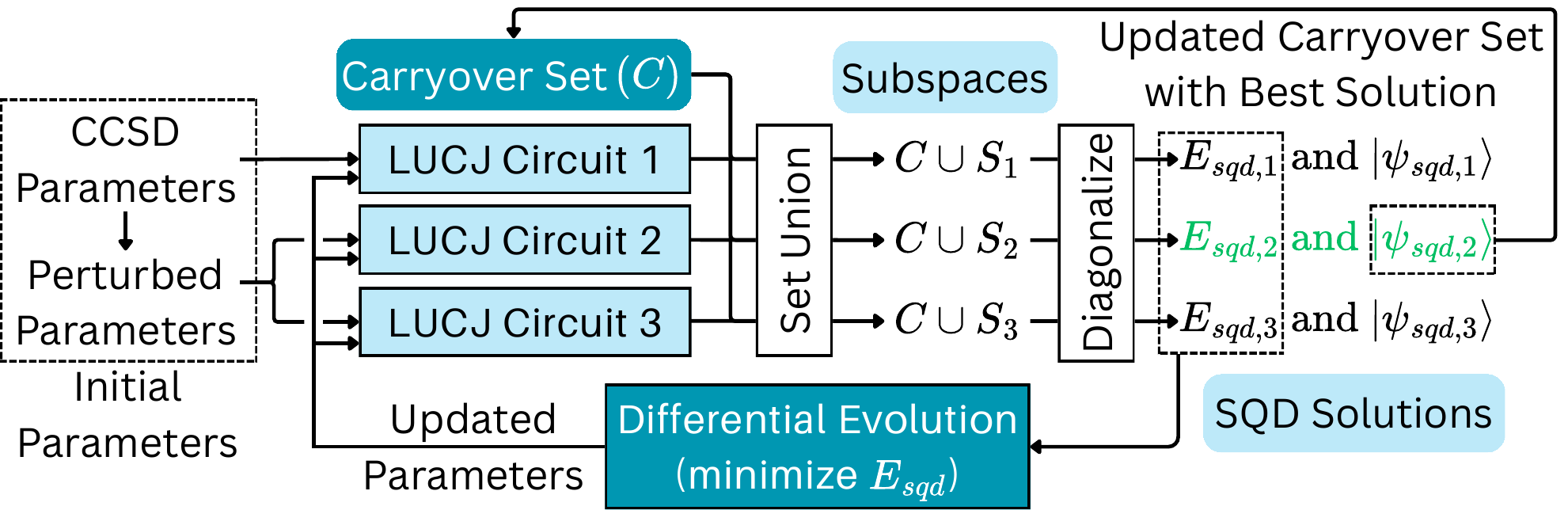}
    \vspace{-6mm}
    \caption{A population of circuit parameters defines multiple LUCJ circuits, which are sampled to generate configurations. Each candidate forms a subspace (combined with a carryover set), whose projected Hamiltonian is diagonalized to obtain the SQD energy used as a fitness value. Differential evolution updates the parameters, while the carryover set retains high-quality configurations across iterations.}
    \vspace{-5mm}
    \label{fig:DE_SQD}
\end{figure}
\begin{figure*}
    \centering
    \includegraphics[width=0.95\linewidth]{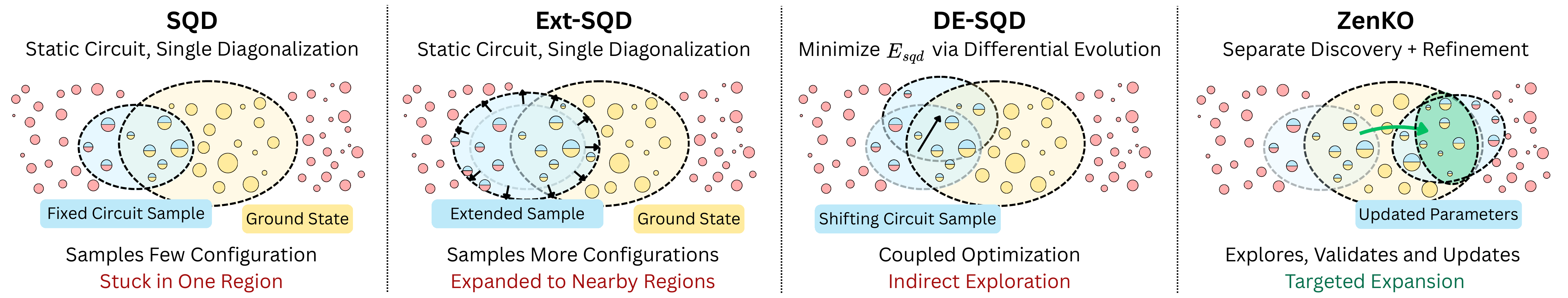}
    \vspace{-2mm}
    \caption{SQD samples a fixed set of configurations and performs a single diagonalization. Ext-SQD expands sampled configurations via excitations but remains dependent on the initial sample. DE-SQD introduces iterative parameter optimization with a carryover set, enabling adaptive refinement but indirect exploration. \sol{} separates discovery and refinement, enabling explicit exploration of new configurations and targeted subspace expansion (more details in Fig.~\ref{fig:zenko}).}
    \vspace{-4mm}
    \label{fig:comparison}
\end{figure*}
\vspace{2mm}
\noindent \textbf{Coupled-cluster Approximation.} Coupled-cluster (CC) methods extend Hartree--Fock by incorporating electron excitations~\cite{hf,ccsd}. Starting from the HF reference, excitation operators promote electrons from occupied to unoccupied orbitals, capturing correlation effects. The resulting nonlinear equations yield an improved ground-state energy. CC methods are categorized according to the level of excitations included. For example, single (S), double (D), and triple (T) excitations correspond to the simultaneous excitation of one, two, and three electrons, respectively. While including higher-order excitations (e.g., triple) can improve accuracy in strongly-correlated systems, it also significantly increases computational cost due to substantially increased parameter/search space.

%potentially 

\vspace{1mm}
In practice, CCSD, which includes single and double excitations, provides a favorable balance between accuracy and computational efficiency~\cite{ccsd_useful}. Recently, CCSD has also been used for a different purpose: providing an initialization for a quantum algorithm that solves electronic spectra~\cite{sqd1,sqd2}.

\begin{figure*}
    \centering
    \includegraphics[width=1\linewidth]{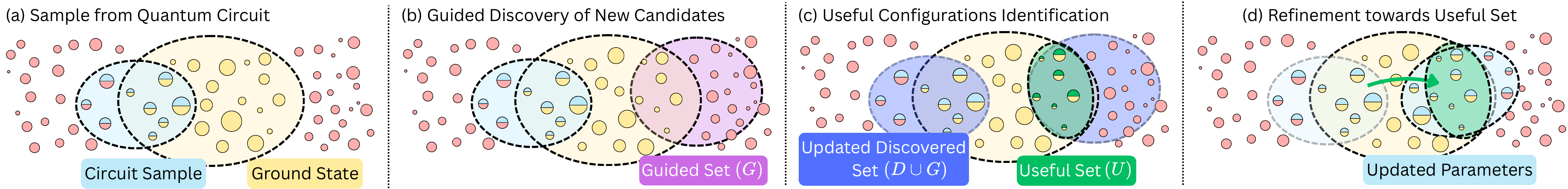}
    \vspace{-6mm}
    \caption{\sol{} iteratively shifts the sampling distribution toward previously missing important configurations and refines the circuit to increase sampling of energetically relevant configurations, providing a new starting point for subsequent iterations. \textbf{(a)} A circuit sample produces configurations with limited overlap with the ground state. \textbf{(b)} Discovery generates a guided set $G_t$ of new configurations by reducing the probability of previously discovered configurations $D_t$, expanding the discovered set as $D_{t+1/2} = D_t \cup G_t$. \textbf{(c)} The expanded set is evaluated using SQD to obtain $|\psi_{\mathrm{SQD},t+1/2}\rangle$, which identifies new configurations that contribute to lowering the energy; those in $G_t$ with non-zero contribution form the useful set $U_t$. \textbf{(d)} The circuit is refined to increase the sampling probability over $U_t$, shifting the sampling distribution towards new useful configurations, enabling subsequent exploration from new regions.}
    \vspace{-4mm}
    \label{fig:zenko}
\end{figure*}

\vspace{2mm}
\noindent \textbf{Sample-based Quantum Diagonalization (SQD).} SQD reduces the combinatorially large Hilbert space to a smaller subspace by focusing on configurations that contribute most to the ground state. While the exact solution requires diagonalization over the full configuration space, in practice, many configurations contribute negligibly. SQD uses a parameterized quantum circuit whose measurement outcomes concentrate on important configurations. In particular, a Local Unitary Cluster Jastrow (LUCJ) circuit~\cite{lucj} initialized with CCSD parameters generates samples that approximate the dominant contributions to the ground state. Each measurement outcome corresponds to a configuration, and these configurations are used to construct a subspace $\mathcal{S} = \{ |\psi_i\rangle, |\psi_j\rangle, \ldots, |\psi_k\rangle \}$. Given this subspace, the full Hamiltonian is projected as

\vspace{-2mm}
\[
\hat{H}_{\mathcal{S}} = \hat{P}_{\mathcal{S}} \hat{H} \hat{P}_{\mathcal{S}}, 
\qquad 
\hat{P}_{\mathcal{S}} = \sum_{\psi \in \mathcal{S}} |\psi\rangle \langle \psi|,
\]\vspace{-2mm}

and then diagonalized classically to obtain the SQD energy and corresponding wavefunction within the constructed subspace. Thus, SQD consists of three steps: (i) sampling configurations from the parametrized circuit, (ii) constructing a subspace from these configurations, and (iii) diagonalizing the projected Hamiltonian, illustrated in Fig.~\ref{fig:background} (right). However, the accuracy of SQD depends critically on the resulting subspace: without relevant configurations, the SQD energy can deviate significantly from the true ground-state energy.

\vspace{2mm}
\noindent \textbf{Challenges and Limitations of SQD.} The primary limitation of SQD is the difficulty of identifying the configurations that contribute to the ground state. Methods such as CCSD are approximate and do not guarantee that all relevant configurations are sampled. A natural approach is to enlarge the subspace by including more configurations, but this reintroduces the original challenge: the cost of diagonalization grows combinatorially. Consequently, the central challenge in SQD is to construct a sufficiently expressive subspace while keeping its size manageable. Next, we review prior approaches that aim to expand the subspace.

\vspace{2mm}
\noindent \textbf{Extended-SQD (Ext-SQD).} The central idea behind Ext-SQD~\cite{ext_sqd} is that excitations of configurations within the current subspace should also make a meaningful contribution to the ground state. Ext-SQD expands configurations sampled from a quantum circuit by applying single and double excitations, generating a significantly larger subspace. This expanded subspace is then passed to a classical diagonalization procedure, which can yield improved estimates of the ground state energy.

\vspace{1mm}

Despite the improvement, the method remains strongly dependent on the initial set of sampled configurations and is inherently static. The method lacks a feedback mechanism to iteratively refine and guide the construction of the subspace toward more relevant regions of the Hilbert space. Next, we explore an optimization-based SQD method.

\vspace{2mm}
\noindent \textbf{Differential Evolution-SQD.} Variational quantum circuits (VQC) are hybrid quantum--classical algorithms that enable iterative optimization of parameterized circuits and are well-suited for near-term quantum devices. In VQC, the cost function is continuous with respect to the parameters, enabling gradient-based optimization methods. In SQD, however, the objective depends on a discrete set of configurations, making it non-smooth with respect to the parameters. Hence, Differential Evolution-SQD (DE-SQD) adopts a population-based stochastic optimization strategy to refine the subspace~\cite{shirakawa2025}.

\vspace{1mm}

DE-SQD begins with an initial population of parameters (CCSD and its perturbations), each defining a circuit that generates configurations, shown in Fig.~\ref{fig:DE_SQD}. These configurations are combined with a carryover set from the previous iteration to form corresponding subspaces, whose projected Hamiltonians are diagonalized to obtain energies. These energies serve as fitness values for the DE algorithm, which updates the parameters, while the best solution is used to update the carryover set. This enables dynamic refinement of the subspace across iterations, with the carryover set retaining high-quality configurations to improve convergence.

\vspace{1mm}

However, exploration of the parameter space remains indirect. The objective is non-smooth with respect to the parameters, so small parameter changes may not alter the sampled configurations or the resulting energies. Consequently, the circuit does not explore the parameter space along well-defined directions, with limited systematic improvement. Additionally, the carryover set biases the method toward previously discovered regions of configuration space, limiting exploration of new configurations that may be critical for improving accuracy. Furthermore, each iteration requires diagonalizing multiple subspaces constructed from the population together with elements from the carryover set. This results in a substantial classical computational cost, particularly as the population size or subspace dimension increases.

\vspace{2mm}
\noindent \textbf{\sol{}: Structured Subspace Discovery and Refinement.}
To address these limitations, we propose \sol{}, a hybrid quantum–classical method that improves both adaptivity and exploration. Unlike Ext-SQD, \sol{} incorporates adaptive learning across iterations. Unlike DE-SQD, it uses gradient-based optimization to discover new subspaces rather than relying on unguided optimization (shown in Fig.~\ref{fig:comparison}).

\vspace{1mm}

\sol{} employs a two-phase iterative procedure. In each iteration, a discovery phase identifies new configurations, which are then evaluated via diagonalization. In the refinement phase, these configurations are used to update the circuit parameters, guiding the search toward new regions of the configuration space. This process is repeated across iterations, enabling progressive exploration and refinement of the subspace while retaining configurations identified in earlier steps. Details of the method are provided in the next section.

\section{\sol{} Design}
\label{sec:design}

Figure~\ref{fig:zenko} summarizes the \sol{} pipeline. The method alternates between two phases: discovery, which expands the sampled subspace to uncover new configurations, and refinement, which concentrates probability on configurations validated by SQD. Discovery uses Hamiltonian-guided coupling to balance exploration and efficiency (Sec.~\ref{sec:discovery}), while refinement employs weighted updates to prioritize configurations with larger contributions to the SQD solution (Sec.~\ref{sec:refinement}). The circuit is parameterized by single- and double-excitation parameters, $T_1$ and $T_2$, which naturally support these phases: $T_2$ enables broader exploration, whereas $T_1$ provides more localized updates for refinement (Sec.~\ref{sec:params}).

\subsection{Guided Discovery of New Configurations}
\label{sec:discovery}

\sol{} expands the sampled subspace to recover configurations not captured by the current circuit (Fig.~\ref{fig:zenko}(a)). We formulate discovery as a gradient-based optimization over circuit parameters that shifts probability mass away from previously discovered configurations.

Specifically, we define the exclusion loss as,
\vspace{-1mm}
\[
\mathcal{L}_{\mathrm{e}}(\theta)
=
\small\sum_{\psi_i \in D_t} p_{\theta}(\psi_i),
\]
\vspace{-1mm}

where $D_t$ denotes the set of configurations discovered up to iteration $t$. We update the circuit parameters $\theta$ using gradient-based optimization to minimize $\mathcal{L}_{\mathrm{e}}$, reducing the probability of sampling known configurations and promoting exploration of new regions. While exclusion promotes exploration, it treats all unseen configurations equally and does not account for their relevance to the target solution. To address this, we guide discovery using information from the current SQD solution. 

Given the current SQD solution ($|\psi_{\mathrm{SQD},t}\rangle$), we incorporate Hamiltonian couplings $\langle \psi_i | H | \psi_{\mathrm{SQD},t} \rangle$ as a proxy for the potential contribution of configuration $\psi_i$. This yields the guided discovery objective
\vspace{-1mm}
\[
\mathcal{L}_{\mathrm{d}}(\theta)
=
\lambda_{\mathrm{e}} \mathcal{L}_{\mathrm{e}}(\theta)
+
\lambda_{\mathrm{g}}
\small\sum_{\psi_i}
\langle \psi_i|  H | \psi_{\mathrm{SQD},t} \rangle \, p_{\theta}(\psi_i),
\]

where $p_{\theta}(\psi_i)$ denotes the sampling probability under the circuit, and $\lambda_{\mathrm{e}}, \lambda_{\mathrm{g}}$ control the balance between exploration and guidance. This objective penalizes revisiting known configurations while prioritizing those strongly coupled to the current SQD solution. The guided-discovery coupling term $\langle \psi_i | H | \psi_{\mathrm{SQD},t} \rangle$  amounts to one Hamiltonian–vector product ($H | \psi_{\mathrm{SQD},t} \rangle$) per outer iteration, followed by an $O(1)$ amplitude lookup. Computed with PySCF's optimized \texttt{contract\_2e} routine, it requires only a fraction of SQD's classical diagonalization time.

\begin{figure}[t]
    \centering
    \includegraphics[width=\linewidth]{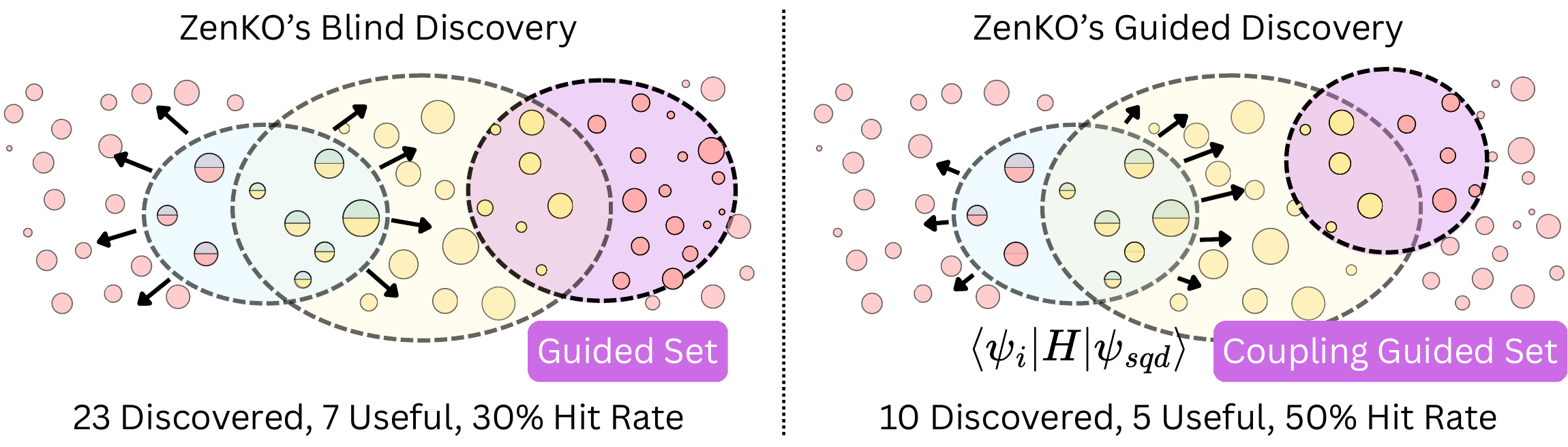}
    \vspace{-6mm}
    \caption{\textbf{Left:} Discovery without coupling explores broadly, producing more configurations but with a lower fraction of useful ones. \textbf{Right:} Guided discovery using $\langle \psi_i | H | \psi_{\mathrm{SQD}} \rangle$ focuses on configurations more likely to lower the energy, yielding fewer but higher-quality configurations, illustrating the trade-off between exploration and efficiency.}
    \vspace{-4mm}
    \label{fig:discovery}
\end{figure}

Guided discovery improves efficiency by focusing on configurations likely to impact the solution, but this guidance is derived from the current SQD solution. As a result, useful configurations that are weakly coupled early on may be underexplored, introducing a trade-off between exploration breadth and efficiency (Fig.~\ref{fig:discovery}), which we control through $\lambda_{\mathrm{e}}$ and $\lambda_{\mathrm{g}}$. In practice, this guided objective provides a favorable balance between subspace growth and computational cost. 

Optimizing either objective produces a sampling distribution that generates a set of new configurations $G_t$ (Fig.~\ref{fig:zenko}(b)). Since $G_t$ contains both relevant and irrelevant configurations, directly updating the circuit would bias it toward suboptimal regions. To avoid this, we discard all parameter updates from discovery and revert to the original parameters $\theta$. Discovery, therefore, acts solely as a subspace-expansion mechanism, without affecting the circuit used for further iterations.

\subsection{Weighted Refinement towards Promising Parameters}
\label{sec:refinement}

We augment the discovered set with the guided set as $D_{t+1/2} = D_t \cup G_t$ and perform SQD on this expanded subspace to obtain an updated solution $|\psi_{\mathrm{SQD},\,t+1/2}\rangle$. 

This solution evaluates the relevance of configurations in the newly explored region. Using it, we define the useful set
\[
U_t = \{ \psi_i \in G_t \; |\; p_{\mathrm{SQD},\,t+\frac{1}{2}}(\psi_i) > 10^{-12}  \},
\]
where $p_{\mathrm{SQD},\,t+\frac{1}{2}}(\psi_i)$ denotes the probability weight of configuration $\psi_i$ in the SQD solution. The set $U_t$ therefore captures the subset of newly discovered configurations that are validated as relevant by the current solution (Fig.~\ref{fig:zenko}(c)).

Although previously discovered configurations are included in the diagonalization, we restrict \( U_t \subset G_t \) to avoid reinforcing already well-represented configurations and instead focus on newly discovered directions. We then define the refinement objective

\[
\mathcal{L}_{\mathrm{r}}(\theta)
=
-\sum_{\psi_i \in U_t} p_{\theta}(\psi_i),
\]
and update the circuit parameters \( \theta \) via gradient-based optimization to increase the sampling probability of configurations in \( U_t \) (Fig.~\ref{fig:zenko}(d)). This concentrates the sampling distribution toward validated regions, whereas discovery expands it toward previously unexplored regions.

\begin{figure}[t]
    \centering
    \includegraphics[width=\linewidth]{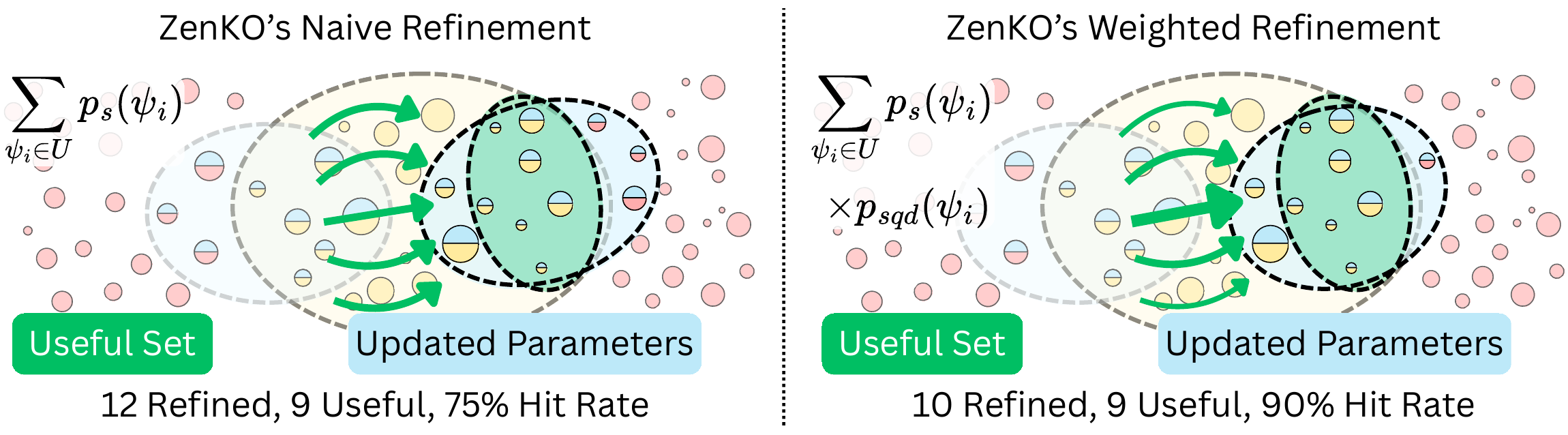}
    \vspace{-6mm}
    \caption{\textbf{Left:} Naive refinement treats all configurations in $U_t$ uniformly, increasing sampling but including less relevant ones. \textbf{Right:} Weighted refinement uses SQD weights to prioritize important configurations, yielding more targeted updates and a higher fraction of useful configurations.}
    \vspace{-3mm}
    \label{fig:refinement}
\end{figure}

While the refinement objective treats all configurations in \( U_t \) equally, their contributions to the SQD solution can differ substantially. To account for this, we weight configurations by their importance in the SQD solution, yielding the weighted refinement objective

\[
\mathcal{L}_{\mathrm{w}}(\theta)
=
-\sum_{\psi_i \in U_t} p_{\theta}(\psi_i)\, p_{\mathrm{SQD},\,t+\frac{1}{2}}(\psi_i).
\]This prioritizes configurations with larger contributions to the solution, focusing updates on the most impactful newly discovered configurations. As a result, the sampling distribution shifts more efficiently toward relevant regions compared to uniform refinement (Fig.~\ref{fig:refinement}). We adopt this as the default strategy due to its improved performance at minimal cost.

Under refinement, parameter updates are retained, reinforcing directions of improvement identified in the current iteration. This shifts the sampling distribution toward newly validated configurations, enabling subsequent iterations to begin exploration from a new, promising parameter space rather than revisiting previously explored ones. In the next section, we examine how different parameter types support these phases.

\subsection{Role of \(T_1\) and \(T_2\) parameterization}
\label{sec:params}

The separation between discovery and refinement is most effective when the circuit parameterization supports qualitatively different types of updates. \sol{} uses CCSD parameters consisting of single-excitation (\(T_1\)) and double-excitation (\(T_2\)) terms, which induce distinct changes in configuration space. 

\begin{figure}
    \centering
    \includegraphics[width=1\linewidth]{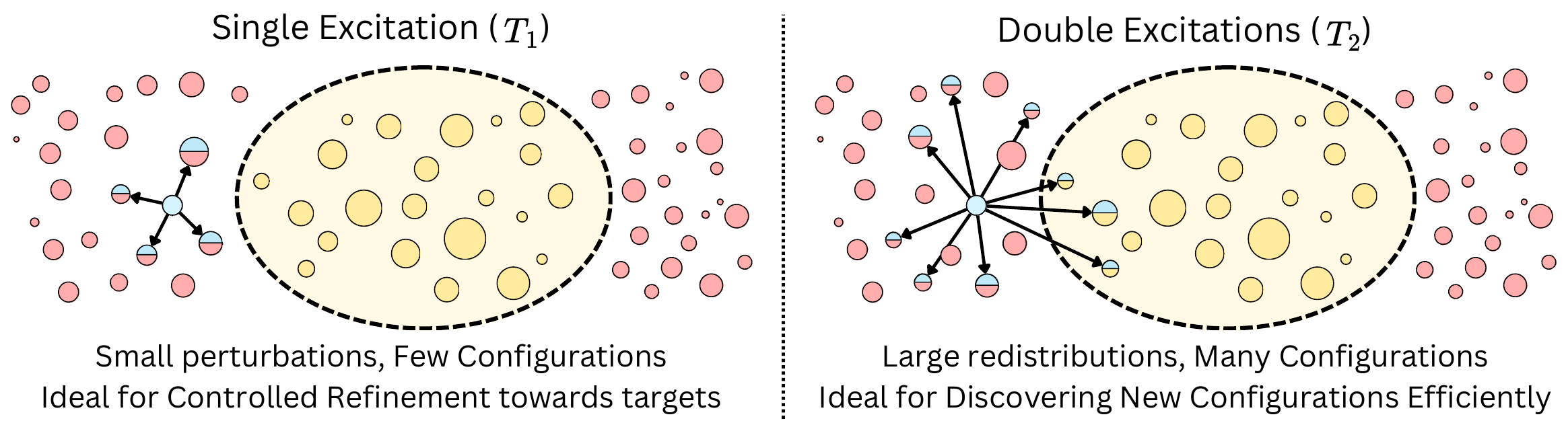}
    \vspace{-6mm}
    \caption{\textbf{Left:} Single excitations ($T_1$) produce fewer, local changes enabling controlled refinement toward target configurations. \textbf{Right:} Double excitations ($T_2$) induce more, nonlocal changes, enabling efficient discovery.}
    \vspace{-4mm}
    \label{fig:param_selection}
\end{figure}

First, \(T_1\) updates tend to connect a configuration to others with more similar occupation patterns than \(T_2\) updates; they only differ by one electron position. Second, the number of possible double-excitations is substantially larger than the number of single excitations, as there are more ways to move two electrons compared to one-electron jumps. Together, these differences cause single-excitations to produce fewer, more local configurations than double-excitations (Fig.~\ref{fig:param_selection}).

This distinction is particularly important for discovery. Because $T_2$ updates induce larger, nonlocal moves, they can reach qualitatively different configurations in fewer steps. In practice, this leads to more selective expansion and discovery of the sampled subspace, avoiding the redundant, local growth associated with $T_1$. As a result, $T_2$ is naturally suited to discovery, where broad exploration is required.

Refinement, on the other hand, primarily requires increasing the sampling probability over a target set of configurations and is therefore less sensitive to the specific parameter subset used. However, we choose $T_1$ for refinement to complement the discovery phase. Assigning \(T_2\) to discovery and \(T_1\) to refinement uses the circuit's degrees of freedom more effectively, enabling efficient exploration followed by targeted redistribution.

In the next section, we present an overview of the iterative algorithm, highlighting the quantum and classical roles of its components within the hybrid framework. 

\subsection{Overview of the full iterative pipeline}
\label{sec:overview}

\sol{} is a hybrid quantum–classical algorithm that alternates between quantum sampling and classical diagonalization. The quantum circuit produces sampled configurations, while classical SQD evaluates them to guide subsequent updates.

Starting from CCSD parameters, circuit samples define an initial subspace, which is classically diagonalized to obtain an SQD solution. This solution guides the discovery phase (quantum), where the $T_2$ parameters of a copied circuit are optimized using a gradient-based objective combining exclusion and Hamiltonian coupling, generating a guided set of new configurations. These updates are applied only to the copied circuit, isolating exploration from the main parameters.

The guided set is merged with previously discovered configurations and re-evaluated using classical SQD to produce an updated solution, which identifies the useful set of configurations that contribute to lowering the energy. In the refinement phase (quantum), the circuit parameters are updated using $T_1$ to increase sampling probability over these configurations, weighted by their SQD contributions.

\begin{figure}
    \centering
    \includegraphics[width=1\linewidth]{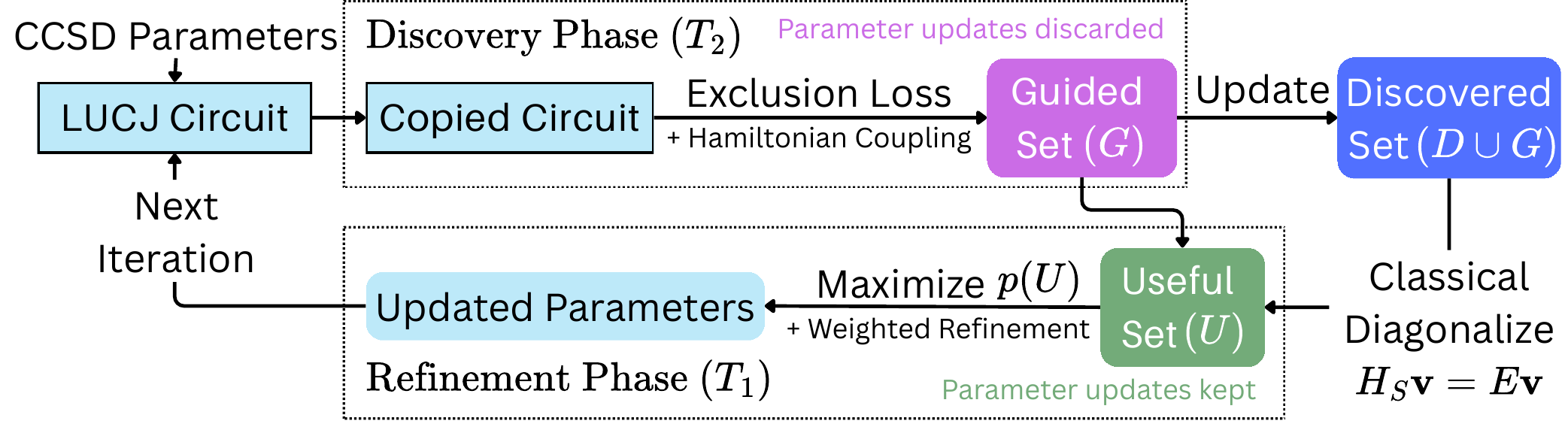}
    \vspace{-6mm}
    \caption{\sol{} alternates quantum discovery and refinement with classical SQD, expanding the configuration space, identifying useful configurations, and reinforcing them to evolve the sampling distribution across iterations.}
    \vspace{-4mm}
    \label{fig:overview}
\end{figure}

These updated parameters define the sampling distribution for the next iteration. Through this loop, the quantum component proposes configurations via sampling and optimization, while the classical component evaluates and filters them via SQD, progressively shifting the sampling distribution toward new promising parameters (Fig.~\ref{fig:overview}).

\section{Methodology}
\label{sec:methodology}
We now describe the experimental setup, including (i) problem instances, (ii) shared SQD pipeline, and (iii) method-specific optimization details.

\vspace{2mm}
\noindent\textbf{Molecular Systems and Active Spaces.} We evaluate \sol{} on molecular systems chosen to probe distinct electronic correlation regimes: weak correlation in small molecules (water, $H_2O$), increasing correlation with system size (hydrogen chains, $H_n$), and strong correlation under bond stretching (molecular nitrogen, $N_2$).

\vspace{3mm}
\noindent\textit{Water ($H_2O$):}
Water is a non-linear molecule with O--H bond length of $0.958 $ \AA \, and an H--O--H angle of approximately $104.5^\circ$. The system has $7$ spatial orbitals and $10$ electrons. Freezing the oxygen $1s$ core orbital, which is chemically inert, yields an active space of $(6o, 8e)$ (spatial orbitals, electrons). Under the Jordan--Wigner transformation~\cite{jordan_wigner}, this corresponds to $12$ qubits and an FCI space of size $\binom{6}{4}^2 = 225$, where the square accounts for independent spin-up and spin-down configurations.

\vspace{1mm}
\noindent\textit{Linear hydrogen chains ($H_n$):}
We consider linear chains of hydrogen atoms with $n = 4, 6, 8, 10, 12$, arranged with a uniform interatomic spacing of $1.0 $ \AA \,. These one-dimensional systems exhibit increasing correlation strength with chain length. Each hydrogen contributes one spatial orbital and one electron, yielding active spaces of $(no, ne)$ with no frozen orbitals. This corresponds to $2n$ qubits and an FCI space of size $\binom{n}{n/2}^2$. For the smallest system ($n = 4$), this yields $8$ qubits and $36$ configurations, while for the largest system ($n = 12$), this yields $24$ qubits and $853,776$ configurations.

\vspace{2mm}
\noindent\textit{Nitrogen ($N_2$):}
We consider the diatomic molecule $N_2$ with $18$ spatial orbitals and $14$ electrons. Freezing one $1s$ core orbital on each atom yields an active space of $(16o, 10e)$. This corresponds to $32$ qubits and an FCI space of size $\binom{16}{5}^2 = 19{,}079{,}424$. To probe both equilibrium and dissociation regimes, we evaluate the system over bond lengths ranging from $0.8 $ \AA \, to $2.5 $ \AA \,, where bond stretching introduces increasingly strong correlation effects. Unless otherwise specified, the case $N_2$ at $1.0$~\AA{} is used as the representative flagship system for detailed analysis and ablation studies.

\vspace{1mm}
These molecular systems are widely used as standard benchmarks~\cite{qsci, te_sqd, ext_sqd}. Hydrogen chains are widely used to study correlated electronic structure~\cite{motta_prx2017}, probing correlations that grow with chain length, and $N_2$ is a standard diatomic benchmark for the same purpose~\cite{qsci, sqd2}. $N_2$ bond stretching spans an approx.\ 19M configuration search space, with \sol{} tracking the dissociation curve far more accurately than SQD, as our evaluation demonstrates in Sec.~\ref{sec:results}. We note that our key ideas and insights are generic and do not exploit the simplicity of any particular molecular structure.

\vspace{1mm}

\begin{figure}
    \centering
    \includegraphics[width=0.9\linewidth]{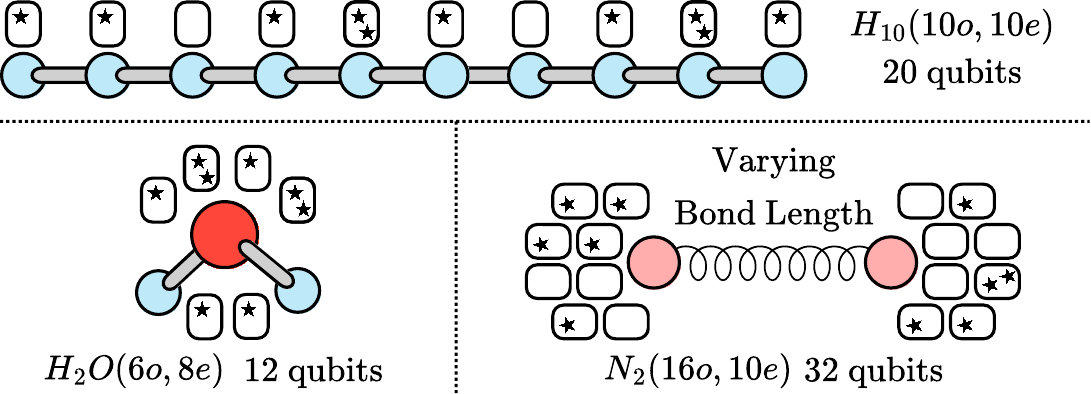}
    \vspace{-3mm}
    \caption{Overview of molecular structures, active spaces ($(o,e)$), and corresponding qubit counts for $H_2O$, $H_{10}$, and $N_2$. Hydrogen chains, $H_n$, with $n = 4 $ to $12$, illustrate scaling ($n=10$ shown for brevity), and $N_2$ probes dissociation via bond stretching.}
    \vspace{-4mm}
    \label{fig:molecules}
\end{figure}

For all systems, molecular Hamiltonians are generated using the PySCF package~\cite{pyscf}. We employ an STO-3G basis for $H_2O$ and hydrogen chains, and a 6-31G basis for $N_2$. All calculations are performed in a spin-restricted formalism. These systems provide a controlled testbed for assessing scalability and robustness across problem sizes and correlation regimes, as summarized in Fig.~\ref{fig:molecules}.

\vspace{2mm}
\noindent\textbf{Classical Reference Methods.} We compute classical reference energies to establish baselines for comparison and to define the ground-truth energy for evaluation. Particularly, we compare against the following methods: Hartree--Fock (HF), CCSD, and full configuration interaction (FCI). HF provides a mean-field baseline, while CCSD captures dynamic correlation through single and double excitations and serves as a strong classical approximation. FCI provides the exact ground-state energy and is used to compute energy error. We next introduce the parameterized quantum circuit used to generate configurations for SQD.

\vspace{1mm}
\noindent\textbf{Quantum Circuit Ansatz.} We employ a spin-balanced UCJ ansatz with a local interaction-pair structure (restricted to nearest-neighbor interactions), following the formulation in Section~\ref{sec:background}. The circuit is applied to the Hartree--Fock reference state and parameterized by single- and double-excitation amplitudes $(T_1, T_2)$. In all experiments, we use a single UCJ layer. The parameters are initialized from CCSD amplitudes, providing a physically motivated initialization that yields configurations with high overlap with the ground state. The ansatz is implemented using the \texttt{ffsim} library~\cite{ffsim}.

\vspace{1mm}
\noindent\textbf{SQD.} We employ SQD following the formulation described in Section~\ref{sec:background}. We use the \texttt{qiskit-addon-sqd} for implementation~\cite{qiskit-addon-sqd}. Circuit samples are generated with 10{,}000 shots. Configurations are filtered using an occupancy tolerance of $10^{-3}$, and the subspace Hamiltonian is diagonalized iteratively (internal SQD optimization) until SQD energy convergence below $0.1$~mHa. A spin constraint of $S^2 = 0$ is enforced, and configurations with amplitudes above $10^{-5}$ are retained within the internal SQD iterations. 

\vspace{3mm}
\noindent\textbf{Evaluation Metrics.}
We evaluate performance using energy error with respect to the FCI ground-state energy and the classical SQD diagonalization time. The ground state energy error is defined as $
\Delta E = E_{\mathrm{SQD}} - E_{\mathrm{FCI}},$ and reported in millihartree (mHa), with chemical accuracy corresponding to an error below 1.6~mHa. Classical SQD diagonalization time measures the wall-clock time required to diagonalize the subspace Hamiltonian. We also report the number of quantum circuit evaluations, the number of discovered configurations, and the dimension of the subspace used in the SQD solver. We report large quantities in K, where K denotes kilo, or $10^3$.

\vspace{1mm}
\noindent\textbf{DE-SQD.} We implement a DE-SQD-style baseline using a population of 5 circuit parameter vectors. One candidate is initialized from CCSD amplitudes, while the remaining candidates are initialized via independent random perturbations of $\pm 0.05$. We adopt the best/2/bin mutation strategy with a mutation factor of 0.25, a carryover ratio of 0.75, and a crossover probability of 0.7, consistent with prior work~\cite{shirakawa2025}. To ensure a consistent optimization budget across systems and avoid premature convergence, we disable tolerance-based stopping by setting $\texttt{de\_tol} = 0$ and run the optimizer for a fixed number of iterations (100 iterations). With a population size of 5, this corresponds to a maximum of 505 function evaluations (sampling and diagonalization) per run. 

\vspace{1mm}
\noindent\textbf{Ext-SQD.}
Ext-SQD augments the configuration set with single and double excitations of each sampled configuration, without additional quantum evaluations~\cite{ext_sqd}. The expanded set is combined with the original configurations and used as input to a single SQD solve, without additional quantum evaluations.

\vspace{2mm}
\noindent\textbf{\sol{}.} Discovery updates the $T_2$ amplitudes for 30 steps with a relatively high learning rate of 0.25, encouraging movement away from previously discovered configurations. An exclusion term with strength $\lambda_e = 10.0$ suppresses rediscovery of known configurations, while a Hamiltonian coupling weight $\lambda_g = 0.3$ introduces a bias toward energetically connected states. Refinement updates the $T_1$ amplitudes for 30 steps with a smaller learning rate of 0.02, as single excitations provide finer-grained control over probability mass redistribution. 

\vspace{1mm}
Each phase employs early stopping based on probability mass. Discovery terminates when the probability mass on previously identified configurations falls below 0.25, indicating successful exploration of new regions. Refinement terminates when the probability mass on high-amplitude configurations exceeds 0.75, indicating successful concentration. The outer loop runs for up to 10 iterations or until SQD energy convergence below $0.1$ mHa
(chosen well below the chemical accuracy at $1.6$ mHa). Both phases use SPSA for gradient estimation with perturbation magnitude $c = 0.1$, requiring two circuit evaluations per update, a total of three, including early stopping evaluation. Parameter updates are performed using Adam with $(\beta_1, \beta_2) = (0.9, 0.999)$ and $\epsilon = 10^{-8}$ to stabilize optimization under stochastic gradients. 

Unless otherwise specified, our reported results (including Figs.~\ref{fig:flag}-~\ref{fig:scaling}) are obtained via ideal quantum simulation of the molecules; this provides us with a better understanding of different components of the \sol{}'s design and their effectiveness, without needing to decouple the stochastic noise effects. Also, we note that \sol{} does not have deeper or larger circuits compared to baseline solutions (in terms of gate count, circuit depth, etc.), and hence, \sol{} is not prone to additional noise effects.

\vspace{2mm}
\noindent\textbf{\sol{}'s robustness under noise model.} We include this evaluation to assess whether \sol{}'s feedback mechanism provides implicit robustness to corrupted samples. The noise model consists of two components. Depolarizing noise replaces a fraction $p_\mathrm{depol} = 0.01$ of sampled configurations with uniformly random bitstrings drawn from the full Hilbert space, which may violate particle-number symmetry. Readout noise independently flips each qubit outcome with probability $p_\mathrm{read} = 0.02$, and can similarly violate particle-number symmetry. We consider four settings: noiseless, depolarizing only, readout only, and both noise channels applied together.

All methods use the same measurement shot budget and SQD solver parameters as \sol{} to ensure a fair comparison and are averaged over 10 seeds for statistical significance.

\section{Results}
\label{sec:results}

\begin{figure}
\centering
\includegraphics[width=1\linewidth]{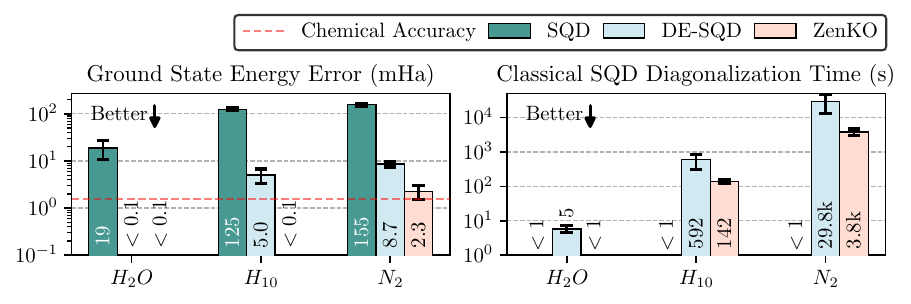}
\vspace{-8mm}
\caption{\textbf{Left:} Ground-state energy error (mHa) for $H_2O$, $H_{10}$, and $N_2$ (lower is better), with the red line denoting chemical accuracy ($1.6$ mHa). \textbf{Right:} Classical SQD diagonalization time (s) (lower is better; results obtained via ideal quantum simulation of the system). \sol{} achieves lower error than DE-SQD while reducing classical diagonalization time, with more gains for larger, strongly correlated systems.}
\vspace{-4mm}
\label{fig:flag}
\end{figure}

We benchmark \sol{} against SQD and DE-SQD on three molecular systems of increasing difficulty: $H_2O$ (bond length $0.958$ \AA, angle $104.5^\circ$), $H_{10}$ (bond length $1.0$ \AA) and $N_2$ (bond length $1.0$ \AA). Hartree--Fock exhibits large errors across all systems (50, 166, and 211 mHa), reflecting its inability to capture correlation, while CCSD performs well in weakly correlated regimes (0.1 and 2.0 mHa for $H_2O$ and $H_{10}$) but degrades for $N_2$ (6.9 mHa), establishing increasing problem difficulty from $H_2O$ to $N_2$.

As shown in Fig.~\ref{fig:flag}, SQD performs poorly, with energy errors of 19, 125, and 155 mHa for $H_2O$, $H_{10}$, and $N_2$, respectively. Although SQD requires less than 1 s of classical diagonalization time, these errors are well above the chemical accuracy threshold of 1.6 mHa and require further augmentation to be practically useful.

DE-SQD employs an iterative optimization algorithm that improves upon SQD, albeit at the cost of substantially greater diagonalization time. In particular, we observe energy errors of 0.1, 5.0, and 8.7 mHa for $H_2O$, $H_{10}$, and $N_2$, respectively. This brings the error for $H_2O$ below the chemical accuracy threshold of 1.6 mHa; however, for the other two molecules, the remaining gap to chemical accuracy is still significant. This improvement is accompanied by a corresponding increase in diagonalization time, which reaches 5 s for $H_2O$, 592 s for $H_{10}$, and a steep 29,800 s for $N_2$, as DE-SQD executes the full SQD pipeline multiple times per iteration.

On the other hand, \sol{} delivers significantly greater improvement than DE-SQD while requiring substantially less diagonalization time. In particular, for $H_2O$ and $H_{10}$, \sol{} achieves near-perfect accuracy with diagonalization times of less than 14\% and 24\% of DE-SQD, respectively. This discrepancy in computational cost becomes even more pronounced for $N_2$, where \sol{} attains an error of 2.3~mHa, substantially closer to the chemical accuracy than DE-SQD's value at 8.7 mHa, while using less than 13\% of the diagonalization time required by DE-SQD.

Across increasingly difficult systems, \sol{}’s advantage remains consistent, simultaneously improving accuracy and reducing computational overhead. In the next section, we analyze the individual components of \sol{} to better understand their contributions.

\subsection{Why it works?}
We begin our analysis by isolating the key components of \sol{}, namely discovery and refinement. Unless otherwise stated, we carry out the remainder of the analysis of \sol{} on the most difficult problem, $N_2$ at a bond length of $1.0\,\text{\AA}$.

\begin{figure}
  \centering
  \includegraphics[width=1\linewidth]{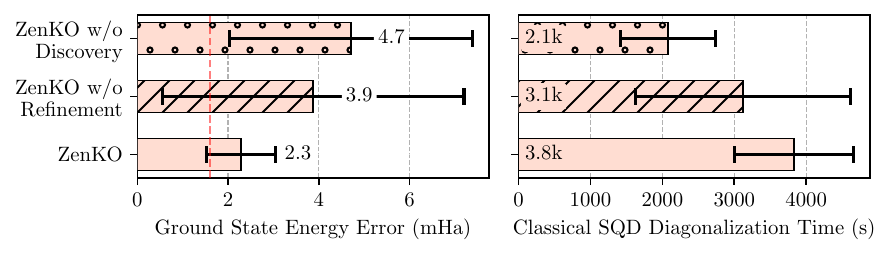}
  \vspace{-6mm}
  \caption{\textbf{Left:} Ground-state energy error for \sol{} and its ablations (without discovery and without refinement), with the red line denoting chemical accuracy ($1.6$ mHa) evaluated on noiseless quantum simulation. \textbf{Right:} SQD diagonalization time. Removing discovery reduces exploration and SQD costs but increases error, while removing refinement slows improvement and increases SQD costs.}
  \vspace{-3mm}
  \label{fig:ablation}
\end{figure}

\vspace{1mm}
\noindent \textbf{Impact of Discovery and Refinement.} In Fig.~\ref{fig:ablation}, we plot the performance of \sol{} against its ablations: \sol{} without discovery and \sol{} without refinement. Each component is disabled by setting its corresponding iteration count to zero.

For \sol{} without discovery (refinement only), we observe the worst energy error at 4.7 mHa, albeit with the lowest diagonalization time. Notably, this performance still surpasses DE-SQD's error at 8.7 mHa. This is partly due to implicit exploration from measurement shots, which yields a limited set of configurations (28.5K vs. 46.7K for full \sol{}), resulting in a smaller subspace and reduced diagonalization cost.

In contrast, \sol{} without refinement (only discovery), shows better performance at an error of 3.9 mHa. In this setting, the circuit explores the configuration space, albeit only around the initialized parameters. While this yields a larger number of discovered configurations (35.0K) and longer diagonalization time, the lack of refinement delays convergence toward high-quality regions of the parameter space. 

Furthermore, the ablations exhibit greater variability in performance. The standard deviation values for the w/o discovery and w/o refinement ablations are 2.67 mHa and 3.33 mHa, respectively, both significantly larger than the 0.76 mHa achieved by the full \sol{} method. Overall, these results establish that both discovery and refinement are essential for achieving stable, accurate performance. Next, we study the impact of the variants of discovery and refinement.

\begin{figure}
    \centering
    \includegraphics[width=1\linewidth]{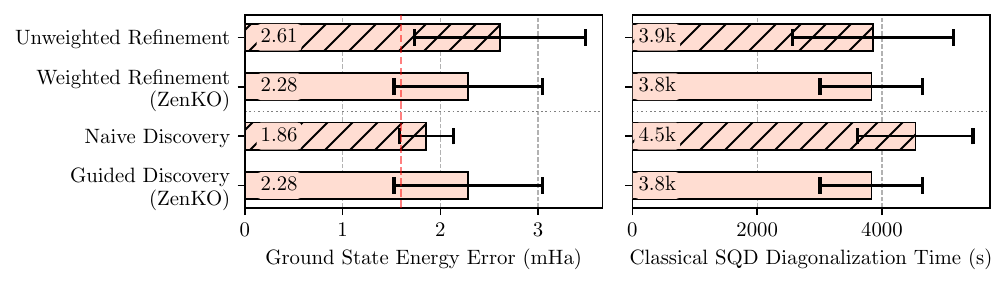}
    \vspace{-6mm}
    \caption{\textbf{Left:} Ground-state energy error for guided vs.\ naive discovery and weighted vs.\ unweighted refinement on ideal simulation, with the red line denoting chemical accuracy ($1.6$ mHa). \textbf{Right:} Guided discovery trades accuracy for lower diagonalization cost, while weighted refinement improves both accuracy and runtime over unweighted refinement.}
    \vspace{-3mm}
    \label{fig:variants}
\end{figure}

\vspace{1mm}
\noindent \textbf{Naive vs.\ Variants.} In Sec.~\ref{sec:design}, we introduced two design choices: naive vs.\ guided discovery, and naive vs.\ weighted refinement. We analyze their impact independently in Fig.~\ref{fig:variants}.

Guided discovery biases the search toward relevant configurations, improving efficiency at the cost of reduced exploration. This tradeoff is reflected in the results: guided discovery achieves 2.28 mHa with 3.8K s, while naive discovery attains lower error (1.86 mHa) but higher cost (4.5K s). For refinement, weighted refinement prioritizes configurations with larger contributions, improving both accuracy and runtime. In contrast, unweighted refinement results in higher error (2.61 mHa) and slightly increased cost (3.9K s).

Overall, guided discovery introduces an accuracy–efficiency tradeoff, whereas weighted refinement improves both accuracy and runtime. In the next section, we study the impact of parameter selection on \sol{}'s performance.

\begin{figure}
  \centering
  \includegraphics[width=1\linewidth]{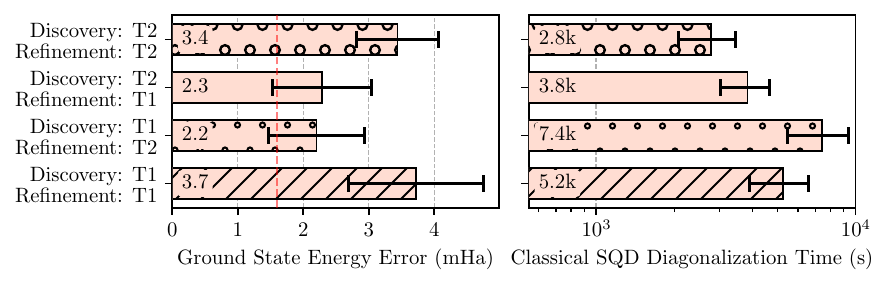}
  \vspace{-6mm}
  \caption{\textbf{Left:} Ground-state energy error for different parameter choices for discovery and refinement to analyze the effect of \sol{} components in reported flagship results in Fig.~\ref{fig:flag} and ~\ref{fig:ablation}, with the red line denoting chemical accuracy ($1.6$ mHa). \textbf{Right:} SQD diagonalization time. Using the same parameters for both phases increases error, while using $T_1$ for discovery increases diagonalization time. The $T_2$/$T_1$ configuration achieves the best accuracy--cost tradeoff.}
  \vspace{-6mm}
  \label{fig:param_grid}
\end{figure}

\vspace{1mm}
\noindent \textbf{Impact of Parameter Choice.} We report the performance of \sol{} for the four possible parameter combinations: discovery/refinement with $T_1$/$T_2$. Recall that the $T_1$ parameters govern fewer, more local interactions, whereas the $T_2$ parameters allow broader, larger numbers of interactions.

As shown in Fig.~\ref{fig:param_grid}(left), the model accuracy is largely independent of whether $T_1$ or $T_2$ is chosen, but rather depends on which combination. In particular, the mixed settings: $T_1$/$T_2$ and $T_2$/$T_1$, achieve errors of 2.2 and 2.3 mHa, respectively. In contrast, the uniform settings: $T_1$/$T_1$ and $T_2$/$T_2$, yield larger errors of 3.7 and 3.4 mHa, respectively. This behavior is expected: using only one parameter regime throughout restricts the search and does not fully exploit the available degrees of freedom.

However, model performance is not determined solely by the energy error. We also plot the classical SQD diagonalization time in Fig.~\ref{fig:param_grid}(right). Here, we observe a sharp increase in runtime whenever discovery uses $T_1$, regardless of whether refinement uses $T_1$ or $T_2$. This supports our intuition that $T_2$, with broader interactions, is better suited for discovery. By contrast, $T_1$ favors local interactions and therefore traverses configuration space gradually, accumulating many intermediate configurations with low contributions along the way (64.7K for $T_1$/$T_2$ vs 46.7K for $T_2$/$T_1$). Overall, parameter choices have a significant impact and must therefore be carefully selected for optimal performance.

\subsection{Comparison with Related Methods}

\sol{} demonstrates superior performance in both accuracy and classical diagonalization time compared to DE-SQD. Next, we provide a detailed comparison with DE-SQD to identify the mechanisms driving these improvements.

\vspace{1mm}
\noindent \textbf{Comparison with DE-SQD.} While \sol{} achieves higher accuracy and lower classical diagonalization time than DE-SQD, it is necessary to assess whether these improvements incur additional quantum resource costs. We evaluate this by comparing the total number of quantum circuit evaluations.

For DE-SQD, the evaluation budget is fixed at 505, derived from 5 populations over 100 iterations, where each iteration performs both sampling and SQD across the full population. This budget is fully utilized by DE-SQD. In contrast, \sol{} employs a hierarchical structure with discovery and refinement phases (30 iterations each) within an outer loop of 10 iterations, yielding a maximum budget of 1800 evaluations. However, \sol{} only uses an average of 573 evaluations.

Therefore, \sol{} reduces classical diagonalization time while maintaining comparable quantum evaluation cost and achieving higher accuracy than DE-SQD. Next, we analyze the evolution of performance during optimization. Since iterations in DE-SQD and \sol{} are not directly comparable, we measure progress as a function of cumulative SQD diagonalization time and report averages across random seeds to account for early convergence in some runs.

\begin{figure}[t]
  \centering
  \includegraphics[width=1\linewidth]{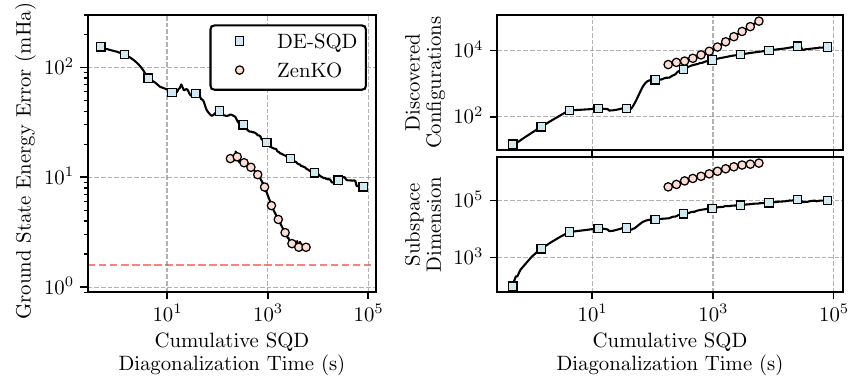}
  \vspace{-6mm}
  \caption{\textbf{Left:} Ground-state energy error of \sol{} and DE-SQD versus cumulative SQD diagonalization time. \textbf{Right:} Configuration set size (top) and SQD subspace dimension (bottom). \sol{} attains lower error with fewer diagonalization costs while maintaining a larger and more diverse subspace.}
  \vspace{-6mm}
  \label{fig:convergence}
\end{figure}

Figure~\ref{fig:convergence} (left) shows the energy error versus cumulative SQD diagonalization time. DE-SQD starts earlier (around $10^{0}$ s), whereas \sol{} begins later (around $10^{2}$ s) due to an initial discovery phase that identifies configurations before the main optimization loop. The initial sampling point of \sol{} is excluded, as it corresponds to a single evaluation and is not directly comparable. Despite this delay, \sol{} converges significantly faster, with a maximum runtime of 5.7K s  compared to 79.4K s for DE-SQD, and consistently achieves lower energy error throughout optimization. To further interpret this behavior, we also examine the number of discovered configurations as a function of cumulative SQD diagonalization time.

For DE-SQD, we use the carryover set as a proxy, as it contains the configurations that most influence the solution and, similar to the discovery set in \sol{}, propagates information across iterations. Figure~\ref{fig:convergence} (right, top) shows that \sol{} consistently discovers more configurations at a given time. At the point where \sol{} first reports values, both methods exhibit comparable configuration counts, a trend that persists up to approximately $10^{3}$ s. Beyond this, \sol{} continues to grow, while DE-SQD plateaus, indicating diminishing returns from further exploration. We next examine the subspace dimension to understand how these configurations are utilized in the subproblem.

The subspace dimension (Fig.~\ref{fig:convergence} right, bottom) follows a different trend. Both methods eventually stabilize, consistent with the plateau in discovered configurations for DE-SQD. For \sol{}, this stabilization is partly due to the SQD solver operating on capped subspaces rather than the full discovered set. Notably, despite similar configuration counts in the early regime, the subspace dimensions differ significantly. This arises from the SQD solver mechanism, which decomposes configurations into components; the number of unique components determines the effective subspace dimension. These results indicate that \sol{} not only discovers more configurations but also yields greater diversity, producing a substantially larger effective subspace.

Overall, \sol{} achieves faster and more accurate convergence by discovering a larger and more diverse set of configurations, leading to an expanded effective subspace.

\vspace{1mm}
\noindent \textbf{Comparison with Ext-SQD.} We also compare \sol{} against Ext-SQD, a fixed expansion constructed from single and double excitations. \sol{} achieves lower energy error ($2.28$ mHa vs.\ $3.61$ mHa). Although Ext-SQD performs worse in terms of accuracy, its fixed procedure requires substantially fewer computational resources. It requires a single quantum evaluation, and the diagonalization step is significantly cheaper, taking approximately $380$ seconds. However, its fixed expansion limits accuracy and prevents further improvement.

We also compare the number of configurations and the subspace dimension. Both methods discover a similar number of configurations ($46{,}667$ vs.\ $43{,}331$), but \sol{} yields a significantly larger subspace ($1827$K vs.\ $583$K). This difference arises because Ext-SQD constructs configurations from samples, leading to higher correlations between the expanded configurations and fewer unique components after decomposition, thereby limiting the SQD subspace.

Comparing \sol{} against DE-SQD, an iterative SQD method, and Ext-SQD, a fixed SQD method, we find \sol{} to be more accurate in both cases. \sol{} is also more efficient than DE-SQD. While \sol{} requires more computational resources than Ext-SQD, the latter remains limited in accuracy due to its fixed expansion strategy and lack of iterative refinement.

\begin{figure}
  \centering
  \includegraphics[width=0.85\linewidth]{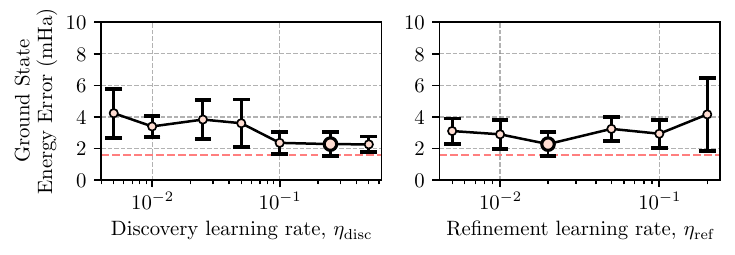}
  \vspace{-4mm}
  \caption{\textbf{Left:} Performance improves with increasing discovery learning rate. \textbf{Right:} Performance degrades with increasing refinement learning rate. The larger markers denote the default values used in the flagship result (Fig.~\ref{fig:flag}).}
  \vspace{-5mm}
  \label{fig:hyperparams}
\end{figure}

\subsection{Effect of Optimization Hyperparameters}

In this section, we analyze the behavior of \sol{} under varying optimization hyperparameters, focusing on the roles of discovery and refinement, in Fig.~\ref{fig:hyperparams}. 

We investigate the impact of the learning rate in both phases. In the discovery phase, higher learning rates lead to improved performance. This can be explained by the objective of discovery: to reduce occupancy on already discovered configurations and promote exploration. With a higher learning rate, the algorithm makes more aggressive updates and can therefore explore a wider range of configurations. In contrast, in the refinement phase, higher learning rates slightly degrade performance. This is consistent with the purpose of refinement, which is to concentrate probability mass on a subset of target configurations. Excessively large updates can cause overshooting, impair stable convergence, and result in suboptimal outcomes.

Finally, we evaluate the impact of shot counts on performance, not shown for brevity. As expected, increasing the number of shots improves performance due to sampling more configurations. However, we observe that the gains begin to plateau beyond $10^4$ shots, with only marginal improvements.

\subsection{Scaling: Bond Length and Molecule Size}
We have considered multiple molecular systems, including $H_2O$, $N_2$, and $H_{10}$, and examined $N_2$ at $1.0\text{\AA}$ in detail. We now focus on two key factors that govern molecular behavior: bond length and size. To isolate these effects, we study $N_2$ at varying bond lengths and hydrogen chains of increasing size.

\begin{figure}
  \centering
    \includegraphics[width=0.99\linewidth]{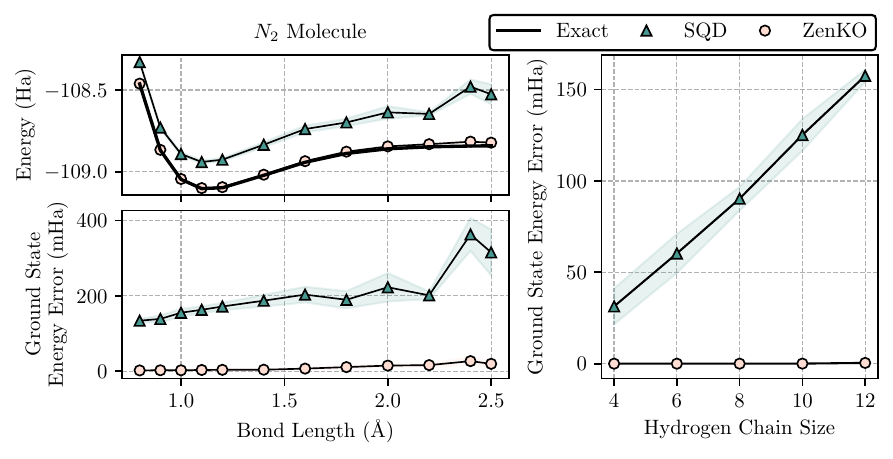}
  \vspace{-7mm}
  \caption{\textbf{Left:} $N_2$ energy and error versus bond length, where SQD degrades at larger separations while \sol{} remains more accurate. \textbf{Right:} Energy error versus system size for hydrogen chains, showing near-linear growth for SQD and consistently low error for \sol{}.}
  \vspace{-6mm}
  \label{fig:scaling}
\end{figure}

In Fig.~\ref{fig:scaling} (left, top), we plot the exact ground state energy for $N_2$ for bond lengths ranging from $0.8\,\text{\AA}$ to $2.5\,\text{\AA}$. Across this range, the exact energy exhibits a characteristic valley shape, as expected, with a minimum near the equilibrium bond length of approximately $1.1\,\text{\AA}$. We observe that SQD by itself performs rather poorly, with a large gap from the exact value exceeding $100\,\text{mHa}$, as shown in Fig.~\ref{fig:scaling} (left, bottom). At longer bond lengths, the interaction is dominated by multiple configurations, making it harder to sample the relevant ones. \sol{} also shows a drop in performance; however, it remains significantly better than SQD. Its error stays well below that of SQD, reaching a maximum of $26.6\,\text{mHa}$, while SQD reaches around $363.2\,\text{mHa}$. 

For the hydrogen chain, we plot the energy error for system sizes ranging from 4 to 12 in Fig.~\ref{fig:scaling} (right). The SQD energy error increases approximately linearly with system size, reflecting steadily increasing electronic correlations as more atoms are added. The error starts from a relatively low value of $31.3\,\text{mHa}$ and grows to $157.5\,\text{mHa}$ as the system size increases from 4 to 12, exhibiting an almost linear trend. \sol{}, on the other hand, maintains near-zero error up to $H_{10}$, with error of $0.4\,\text{mHa}$ at $H_{12}$. Compared to SQD, these values are significantly closer to the ground state energy. Overall, \sol{} maintains strong performance across both scaling regimes. Next, we examine its performance with noise.

\subsection{Robustness against Noise}
\begin{figure}
    \centering
    \includegraphics[width=1\linewidth]{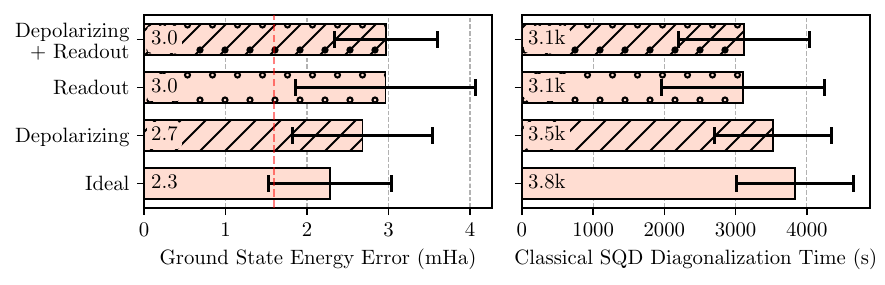}
    \vspace{-7mm}
    \caption{\textbf{Left:} Ground-state energy error of \sol{} under noiseless, depolarizing, readout, and the combined noise-model's quantum simulations, showing only modest degradation.  \textbf{Right:} SQD diagonalization time across noise settings remains comparable.
    }
    \vspace{-6mm}
    \label{fig:noise}
\end{figure}

We examine the robustness of \sol{} under noisy sampling (Fig.~\ref{fig:noise}), considering depolarizing noise, readout noise, and their combination (Sec.~\ref{sec:methodology}).

\sol{} maintains consistent performance across all noise settings, with the worst energy error reaching $3.0\,\text{mHa}$ compared to $2.3\,\text{mHa}$ in the ideal case. The largest degradation is observed under readout noise. This degradation in energy accuracy is accompanied by a reduction in SQD runtime, with runtime decreasing to 3.1K s in the worst-case error. This behavior is expected. Depolarizing noise is implemented using uniformly random samples that contribute little useful information and can be effectively filtered out, reducing the effective search space with limited impact on accuracy. In contrast, readout noise corrupts valid samples, making optimization more difficult and degrading performance.

Overall, these results indicate that our iterative gradient-based optimizer, \sol{}, is robust against moderate noise.

\section{Related Work}
\label{sec:rel_works}

Quantum-selected configuration interaction (QSCI), also known as sample-based quantum diagonalization (SQD), constructs configuration subspaces from measurements of parameterized quantum circuits~\cite{qsci, ext_sqd, te_sqd}. Early approaches rely on classically prepared reference states, such as coupled-cluster wavefunctions or variationally optimized circuits, to define the sampling distribution~\cite{sqd1, sugisaki2025hamiltonian}. To reduce this dependence, time-evolved QSCI (TE-QSCI) generates configurations by sampling from Hamiltonian time-evolved states, enabling subspace construction without classical pre-optimization~\cite{te_sqd, sugisaki2025hamiltonian}. Extensions such as Ext-SQD improve subspace completeness by augmenting sampled configurations with their excitations~\cite{ext_sqd}, while population-based methods like DE-SQD introduce iterative updates to circuit parameters~\cite{shirakawa2025}.

Recent work identifies a key limitation of QSCI/SQD: the configuration subspace is determined by a fixed sampling distribution, resulting in repeated sampling of previously observed configurations and limiting the discovery of new configurations, thereby leading to poor performance~\cite{sqd_limitation}. Adaptive approaches such as ADAPT-QSCI iteratively improve the input state by selecting operators that reduce the energy~\cite{adapt_qsci}. Similarly, population-based methods such as DE-SQD update circuit parameters across iterations~\cite{shirakawa2025}. However, these updates remain energy-driven and do not explicitly promote the discovery of new configurations beyond those already supported by the current state. As a result, improvements may become less efficient, with optimization increasingly favoring configurations already represented in the state.

These limitations indicate that effective subspace construction requires iterative, guided updates to the sampling process rather than passive or purely energy-driven refinement. This motivated \sol{}, which introduced an explicit separation between configuration discovery and refinement within an iterative SQD framework.

\section{Conclusion}
\label{sec:conclusion}

Sample-based quantum diagonalization faces a fundamental bottleneck: achieving better accuracy requires sampling more configurations, but a larger subspace incurs higher classical diagonalization cost. \sol{} addresses this through systematic exploration, decomposing configuration selection into guided discovery, SQD-based validation, and weighted refinement, with circuit parameters specialized to each phase: $T_2$ for discovery and $T_1$ for refinement.

\sol{} achieves higher ground-state accuracy than other SQD-based methods, at a fraction of the diagonalization cost of the prior iterative approach. These improvements hold across noise models, $N_2$'s bond-length range, and hydrogen chains of increasing size, highlighting \sol{}'s robustness.

\section*{Acknowledgment}
We thank the anonymous reviewers for their constructive feedback, which helped improve this work. This research was supported in part by Northeastern University and NSF Award 2144540. We used Claude, an AI language model developed by Anthropic, for limited writing assistance; all such text was reviewed and verified for correctness by the authors.

\bibliographystyle{IEEEtran}
\bibliography{references}

\end{document}